\documentclass[nonacm, sigconf]{acmart}

\setcopyright{none}
\renewcommand\footnotetextcopyrightpermission[1]{} % removes footnote with conference information in first column
\AtBeginDocument{%
  }

\acmConference[]{}{}{}
\usepackage{graphicx}

\usepackage{listings}
\usepackage{xcolor}
\begin{document}
\begin{sloppypar}

\title[]{Shotit: compute-efficient image-to-video
search engine for the cloud}


\author{Leslie Wong}
\affiliation{%
  \institution{}
  \city{Shenzhen}
  \country{China}}
\email{lesliewong1@acm.org}

\renewcommand{\shortauthors}{Leslie Wong}

\begin{abstract}
With the rapid growth of information technology, users are exposed to a massive amount of data online, including image, music, and video. This has led to strong needs to provide effective corresponsive search services such as image, music, and video search services. Most of them are operated based on keywords, namely using keywords to find related image, music, and video. Additionally, there are image-to-image search services that enable users to find similar images using one input image. Given that videos are essentially composed of image frames, then similar videos can be searched by one input image or screenshot. We want to target this scenario and provide an efficient method and implementation in this paper.

We present Shotit, a cloud-native image-to-video search engine that tailors this search scenario in a compute-efficient approach. One main limitation faced in this scenario is the scale of its dataset. A typical image-to-image search engine only handles one-to-one relationships, colloquially, one image corresponds to another single image. But image-to-video proliferates. Take a 24-min length video as an example, it will generate roughly 20,000 image frames. As the number of videos grows, the scale of the dataset explodes exponentially. In this case, a compute-efficient approach ought to be considered, and the system design should cater to the cloud-native trend. Choosing an emerging technology - vector database as its backbone, Shotit fits these two metrics performantly. Experiments for two different datasets, a 50 thousand-scale Blender Open Movie dataset, and a 50 million-scale proprietary TV genre dataset at a 4 Core 32GB RAM Intel Xeon Gold 6271C cloud machine with object storage reveal the effectiveness of Shotit. A demo regarding the Blender Open Movie dataset is illustrated within this paper. For source code of Shotit, please refer to \url{https://www.github.com/shotit/shotit/}.
\end{abstract}



\begin{teaserfigure}
  \includegraphics[width=\textwidth]{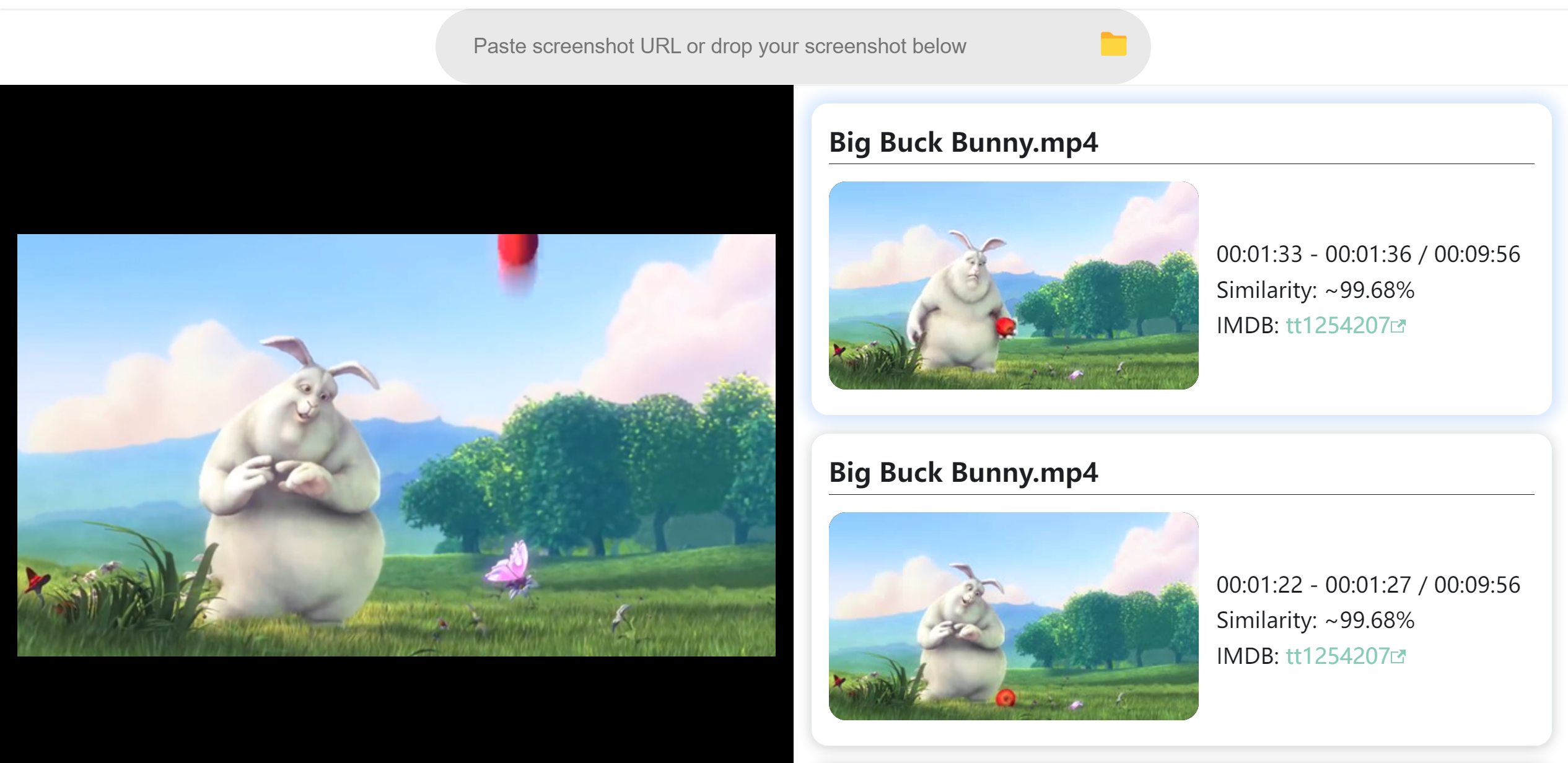}
  \caption{Shotit Demo Regarding Blender Open Movie}
  \Description{Shotit Demo Regarding Blender Open Movie}
  \label{fig:teaser}
\end{teaserfigure}


\maketitle

\section{Introduction}
With the rapid development of Machine Learning/Deep Learning, researchers in the domain of computer vision apparently spend more and more attention to solving computer vision problems utilizing convolutional neural network(CNN) models, such as ResNet50\cite{resnet50}, Xception\cite{xception01}, VGG16\cite{vgg16}, MobileNetV3Large\cite{mobilenetv3}. However, these models share an implicit pitfall, that is their memory usage is tremendously high. Specifically, when it comes to the image-to-image search problem, the approach that a machine learning engineer typically take is to use the penultimate layer of these CNN models\cite{CNNintro01} to generate an image's vector information and search the similar vectors to match its similar images. 

Coming to the scenario of image-to-video search, which is essentially image-to-image search, it is not appropriate to take the machine learning approach when it comes to low-compute limitation. Searching around the web, there exists a particular solution trace.moe\cite{tracemoe01} that resolves image-to-video by using the visual descriptor information of images to match the frames of similar videos so as to find videos. It classifies itself as content-based image retrieval(CBIR)\cite{cbir01}. Since trace.moe is open-sourced, we looked through its source code and found the descriptor used is Color Layout\cite{colorlayout01}, which captures an image's dominant color information in equally sized sub-images via 8 * 8 girds, with no ML models needed, exemplified as Figure. 2 and Figure. 3. Its functionality to index and search Color Layout hashes is powered by LireSolr\cite{liresolr01}, a solr plugin to Apache Solr\cite{solr01}. LireSolr is an implementation of CBIR in Java. Combining OpenCV\cite{opencv01} and ffmpeg\cite{ffmpeg01}, trace.moe provides full-fledged functionality of image-to-video search in its particular genre - anime. The search experience of its running website is decent.

\begin{figure}[h]
  \centering
  \includegraphics[width=\linewidth]{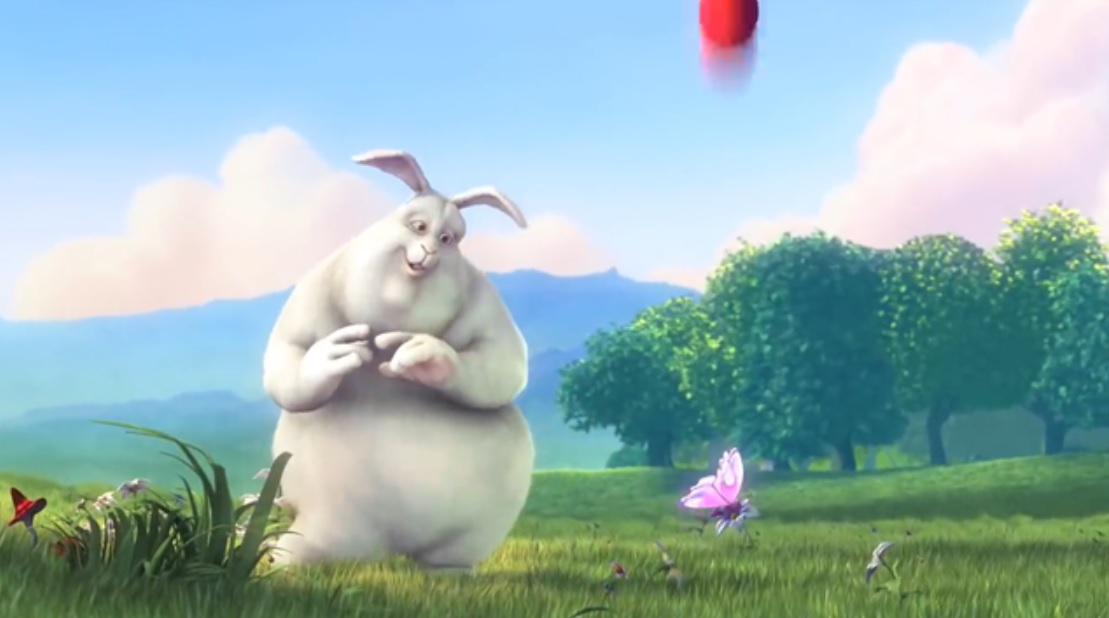}
  \caption{Original Shotit Demo Image}
  \Description{Original Shotit Demo Image}
\end{figure}

\begin{figure}[h]
  \centering
  \includegraphics[width=\linewidth]{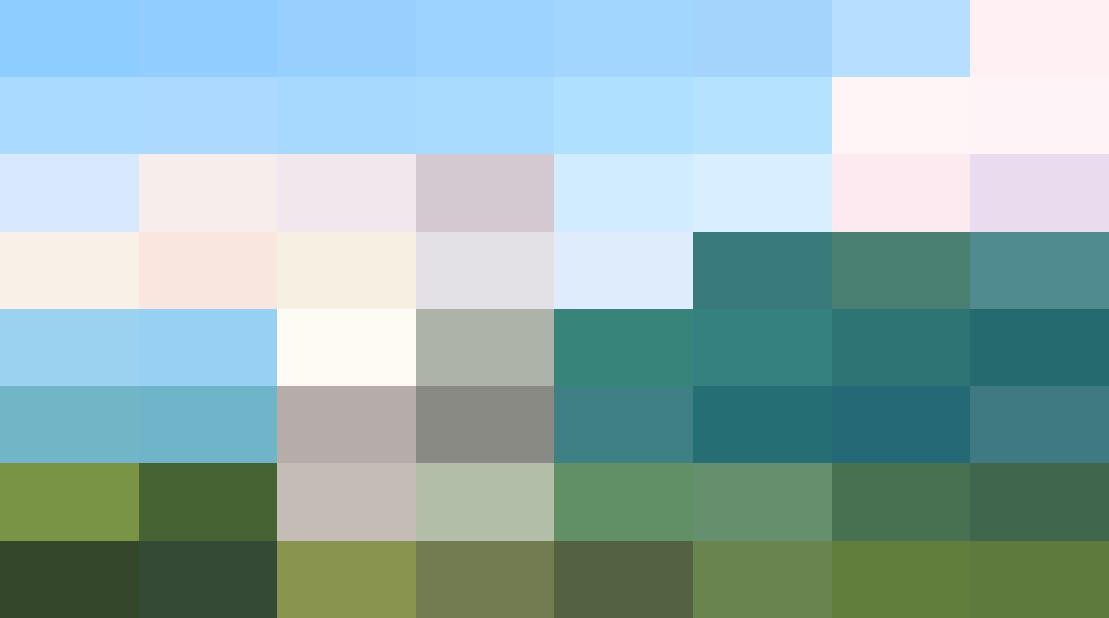}
  \caption{Processed Shotit Demo Image}
  \Description{Processed Shotit Demo Image}
\end{figure}

However, when we managed to pull and run trace.moe locally, the limitation of it was disclosed. The index procedure of trace.moe is as follows\cite{tracemoeslide01}. Raw Video -> FFmpeg extract all frames -> Lire extract image features -> deduplicate images with same image features with a running window of 12 frames -> append timestamp -> load into solr. And the search operation is to directly search from LireSolr, optionally cutting the black borders of the target image using OpenCV. The pitfall focuses on Apache Solr, a full-text search engine built on Apache lucene\cite{lucene01}.

Due to the plugin mechanism of LireSolr, trace.moe is designed as single-machine mode\cite{tracemoe2018}, not SolrCloud\cite{solrcloud01} recommended by the Apache Solr official. It splits up its hash data into 32 solr cores and performs search concurrently with one high-end multi-core machine, (2 x E5-2696v4 (44 Cores)), 512GB RAM, 3 x 16 TB HDD, 10G LAN\cite{tracemoeslide02}. The volume of trace.moe's dataset to search is around 8 hundred millions\cite{tracemoeabout01}, and when searching it will reduce the search volume by Locality Sensitive Hashing\cite{lsh01}. Users of trace.moe can search for decent results within seconds.

Turning back to our local deployment, such metrics are hard to achieve. Apache Solr uses JVM. Each solr core represents one JVM. As the volume of dataset increases exponentially, the reality that LireSolr needs to use JVM to load hash into memory continuously will lead to slower search speed. As the author of LireSolr explained in his book\cite{lirebook01}, {\itshape LIRE serves as a good example for linear search in Java. In LIRE, linear search is the default approach, mainly to reduce complexity of usage for novice users, but also due to its satisfactory performance for small and moderate-size image repositories.} Because of this, the author of trace.moe set up a high-end machine to satisfy the hardware need of its dataset. 

Given that memory is much faster than disk, wouldn't in-memory computing be a good alternative when it comes to search? For example, Apache Spark\cite{spark01} follows the philosophy of in-memory computing and performs excellently in large-scale data analytics. With this intuition, relevant literature regarding search was searched and analyzed. Searching related information about image-to-image search, an emerging technology called approximate nearest neighbor(ANN) search\cite{anns01} was noticed.

The term {\itshape nearest neighbor search} occurs on the LireSolr book\cite{lirebook01} as well. However, the context is different. As the development of ML/DL emerges rapidly, vector-based data plays an increasingly important role. Researchers and developers in the database area commence developing a new kind of database dedicated to vector-based data, namely vector database. Under the hood, it is the approximate nearest neighbor search. Faiss\cite{faiss01}, the full name of which is Facebook AI Similarity Search, is a performant open source implementation to approximate nearest neighbor search. It is a library for efficient similarity search and clustering of dense vectors. Centering on Faiss, a significant number of vector databases arouse, whose functionality is like ANN + SQL\cite{anns01}, e.g, Pinecore\cite{pinecone01}, Qdrant\cite{qdrant01}, Vald\cite{vald01}, OpenDistro\cite{opendistro01}, Milvus\cite{milvus01}, Vsearch\cite{vsearch01}, Weaviate\cite{weaviate01}. Among them, Milvus satisfies the desire to refactor trace.moe. It provides a nodejs sdk. Previously implemented in PHP, the latest version of trace.moe is written in JavaScript. Hence it comes the intuition that Milvus could be integrated into trace.moe to speed up search. 

\begin{figure}[h]
  \centering
  \includegraphics[width=\linewidth]{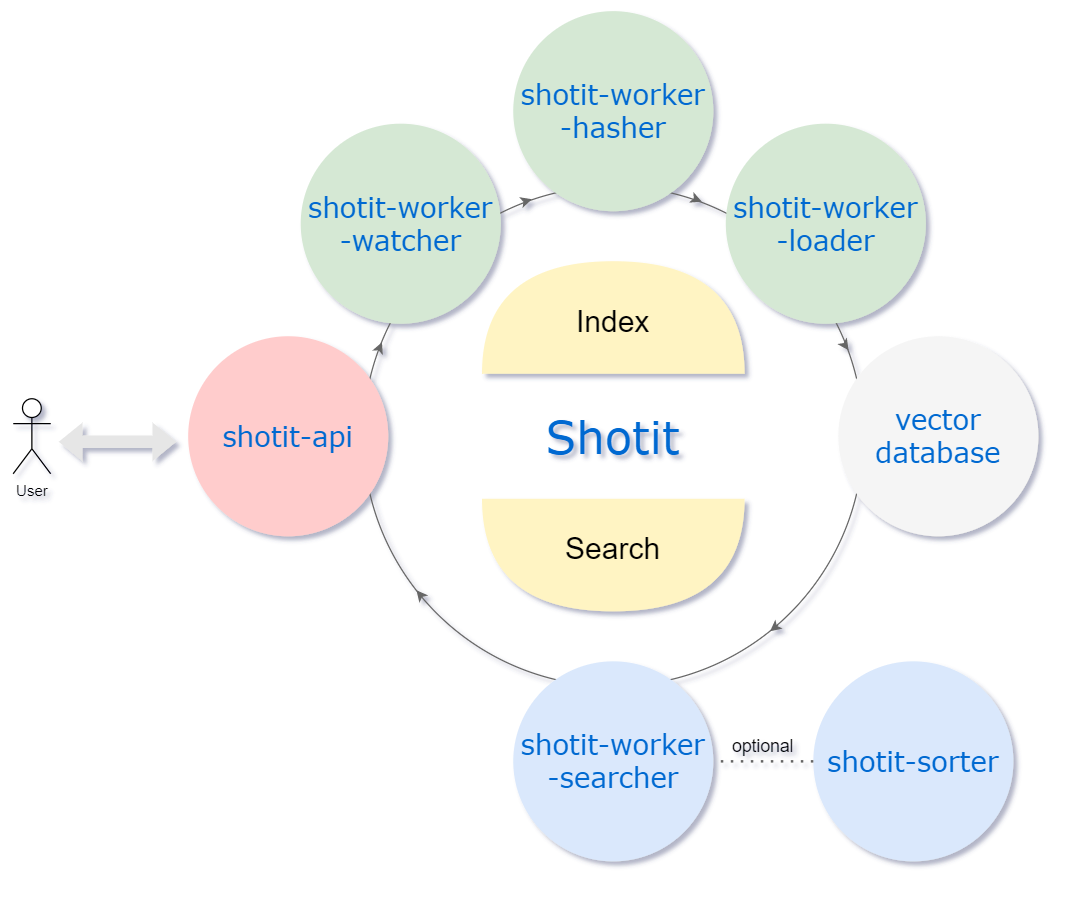}
  \caption{Brief system architecture of Shotit}
  \Description{Brief system architecture of Shotit}
\end{figure}

Figure.4 demonstrates the brief architecture of Shotit. It leverages the index infrastructure of trace.moe and chooses the vector database Milvus instead as its backbone to power search. The whole procedure of Shotit is as follows. When indexing, a massive amount of videos are uploaded to a folder that the shotit-worker-watcher keeps detecting. When shotit-worker-watcher detects the upload, it will upload the videos to object storage, locally or remotely. Then shotit-worker-hasher will pull the videos down and use LireSolr to generate hash data as compressed XML files. And shotit-worker-loader will unzip those XML files, get the hashes, convert them into vectors, and insert them into the vector database. When searching, the shotit-worker-searcher will act as a delegate to search the designated results from the vector database by comparing cosine distance, and optionally the results can be reranked by shotit-sorter, a Keras\cite{keras01}/Faiss powered middleware, to increase the correctness of the Top1 result. Both the index and search procedure are monitored by shotit-api. Users could receive JSON data containing image and video clip links from the restful api that shotit-api provides. 

One key factor that Shotit differentiates its performance from the CNN model based system is the dimension of its Color Layout hash vector. As Table.1 shows, the dimension of the vector generated by Color Layout is only 100, compared with MobileNetV3Large's 1000, Xception's 2048, ResNet50's 2048, and VGG16's 4096. 

When scaling to a million-scale or even billion-scale dataset, this advantage benefits significantly. The in-memory computing mechanism of vector databases requires a lot of memory. The nature of Color Layout's low dimension can mitigate the proliferation of memory use while retaining satisfactory correctness. Since the development of vector databases is emerging and highly optimized for search performance. Combining the advantage of Color Layout's low dimension and vector database's search capability boosts the search performance of image-to-video tremendously. The vision of Shotit is to make image-to-video search engines genre-neutral, ease-of-use, compute-efficient, and blazing-fast.

\begin{table}
  \caption{Comparison of vector dimension}
  % \label{tab:freq}
  \begin{tabular}{ccl}
    \toprule
    Image descriptor or ML model&Vector dimension\\
    \midrule
    Color Layout & 100\\
    MobileNetV3Large & 1000\\
    ResNet50 & 2048\\
    Xception & 2048\\
    VGG16 & 4096\\
  \bottomrule
\end{tabular}
\end{table}

In the remainder of this paper, RESEARCH BACKGROUND elucidates some related works of image-to-video search, some inspirations from other domains such as cloud-native and music retrieval that could facilitate image-to-video search, and a brief overview of ANN search. THEORY REASONING elaborates two typical approaches to image-to-video search, the CNN approach and the LireSolr approach proposed by trace.moe. Pros and cons of them are discussed and the optimized approach that Shotit takes is introduced. ARCHITECTURE DESIGN illustrates the big picture of Shotit in detail about how it applies to local standalone deployment and cloud-native distributed deployment. NOTABLE OPTIMIZATION POINTS provides explanations of some optimization works that we inherit from trace.moe as well as some we tackle on our own. PERFORMANCE BENCHMARKS demonstrates the experiments we performed with two datasets, the Blender Open Movie dataset, and a proprietary TV genre dataset and explained its performance progress from a numerical point of view. In CONCLUSION \& FUTURE WORKS we would conclude our contributions and come up with some optimization directions that might be of help to the future development of Shotit. 

\section{RESEARCH BACKGROUND}
In this section, we elucidate some related works in the following domains which benefit developing Shotit, image-to-video search, cloud-native technologies, music retrieval, peer-to-peer file sharing, and approximate nearest neighbor search. 

\textbf{Image-to-video search}. The research of image-to-video search has a long history. From the Color Layout paper\cite{colorlayout01} published in 2001, we learned the researchers revealed that they used the Color Layout descriptor to search over 24 hours of videos in less than a second. However, the hardware requirement for their success is agnostic. 

The typical solution to image-to-video search is to take each frame in the video as a separate image, such that it could be resolved as to whether the target image is similar to certain image frames. Shotit does take this approach fundamentally. However, one obvious problem is that it would contain many duplicated image frames. To prevent this, Shotit reduces duplicated image frames by using the Color Layout hash to compare the near exact image frames and squash them into only one. This would be more intuitive in the following ARCHITECTURE DESIGN section. 

On the other hand, video-clip level retrieval is investigated by some researchers\cite{videoclip01}. After the initial video-clip level retrieval, a frame-level inspection is performed for the most promising video clips. 

When it comes to the technique to compare and retrieve, the classic one is to use image descriptors of global features or local features. Take LireSolr\cite{liresolr01} as an example, it has implemented twelve kinds of global features, PHOG(pyramid histogram of oriented gradients), Opponent Histogram(simple color histogram in the opponent color space), Color Layout(from MEPG-7), Scalable Color(from MEPG-7), Edge Histogram(from MPEG-7), CEDD(very compact and accurate joint descriptor), FCTH(more accurate, less compact than CEDD), JCD(joined descriptor of CEDD and FCTH), Auto Color Correlogram(color to color correlation histogram), SPCEDD(pyramid histogram of CEDD), Fuzzy Opponent Histogram(fuzzy color histogram), Generic Global Short Feature(generic feature used to search for deep features in LireSolr). One other technique is to use a pre-trained convolutional neural network to generate features, as portrayed previously, the dimension of which is quite high. Two works of literature we found provide implementation utilizing this technique\cite{CNNI2V1024}\cite{CNNI2V4096}, the first of which is 1024 dimensional vector and the second is 4096. 

Shotit chooses the Color Layout hash of LireSolr to index video data because Color Layout requires low hardware requirements even without GPU and has been tested production-grade by trace.moe for years. Most importantly, the Color Layout hash of LireSolr can be converted to a 100-dimensional vector to insert into the search-performant vector database. The experiment we performed with a proprietary TV genre dataset is satisfactory. We would provide detailed reasoning about this in the THEORY REASONING section.

\textbf{Cloud-native technologies}. Before introducing cloud-native, big data should be discussed to supplement the context. The concept of big data is ignited by three influential papers published by Google, Google File System\cite{GFS01}, MapReduce\cite{MapReduce01}, and Big Table\cite{BigTable01}. Inspired by Google's big data implementation, Hadoop\cite{hadoop01} provides its respective open-source implementation, HDFS\cite{hdfs01}, Hadoop MapReduce\cite{HadoopMapReduce01}, and HBase\cite{hbase01}. Due to the network IO limitation at that time, the whole distributed system of Hadoop is composed of numerous homomorphic host computers, with compute unit and storage unit bundled together. When new resources are needed. adding new homomorphic host computers is the solution.

Then, the era of cloud computing came. The strategy that AWS takes at designing the cloud availability zone is to allow EC2 instances to access S3 object storage within the same zone freely and swiftly\cite{aws01}. Other cloud computing vendors follow this convention, providing an opportunity to redesign the big data distributed system to separate compute unit and storage unit under the cloud environment. Among such practices, Snowflake stands out significantly\cite{snowflake01}. Its shared-storage architecture is widely recognized and adopted.

To achieve the goal of compute-efficient for Shotit, the shared-storage architecture is reasonable. We will elaborate the specific paradigm we take in the ARCHITECTURE DESIGN section.

\textbf{Music retrieval}. The mindset of music retrieval is quite similar to image-to-video search. Users provide a seconds-long music clip and want to know the music track title and artist. The research milestone in this area is the Shazam algorithm\cite{shazam01}, which identifies the key to audio fingerprinting. Given Shazam receives great commercial success\cite{shazam02}, we believe the development of Shotit is prospective.

\textbf{Peer-to-peer file sharing (P2P)}. Renowned for its decentralized manner, peer-to-peer file sharing enables users to transfer files over the internet. The underlying file sharing protocol is called BitTorrent\cite{BitTorrent01}. Two important concepts utilized by P2P are worth mentioning. The first is to take advantage of the upload bandwidth of joint distributed micro servers from phone or PC. The second is to split up the file being distributed into tiny segments, whose hash information is shared across the distributed system to fetch and compare. 

Since the purpose of Shotit is to use image to retrieve the correct video clip and information, video processing takes up an important segment of the Shotit architecture. Designed as cloud-native, Shotit needs to fetch the video files from the storage unit to the compute unit to generate video clips for users. Fetching the video files as a whole and then generating clips would cause significant network IO impact. So a better way is to divide the video files first. P2P's file splitting idea gives us a direction. The specific implementation detail is disclosed in the ARCHITECTURE DESIGN section.

\textbf{Approximate Nearest Neighbor Search}. Reviewing Approximate Nearest Neighbor Search(ANNS), IEEE Conference on Computer Vision and Pattern Recognition(CVPR) 2020 organized a tutorial session entitled {\itshape Image Retrieval in the Wild}, and among the presentations the report {\itshape Billion-scale Approximate Nearest Neighbor Search} provides a comprehensive review of this topic\cite{anns01}. The leading actor in this report is Faiss\cite{faiss01}. It discusses Nearest Neighbor Search and Approximate Nearest Neighbor Search as well as extrapolates the state-of-the-art implementation of ANN in Python as of 2020.

Despite its mathematical inference about the algorithms and benchmarks, the most enlightening point to our development of Shotit is that it mentions several industrial-strength nearest neighbor search engines at the end of the slide\cite{anns02}, according to its words, "something like ANN + SQL", Vsearch\cite{vsearch01}, OpenDistro\cite{opendistro01}, Milvus\cite{milvus01}, Vald\cite{vald01}. After surveying them, we found Milvus fits our needs the most. As mentioned before, it provides a nodejs sdk to match trace.moe's JavaScript codebase almost seamlessly. Besides, it is open sourced and the community of it is flourishing. Since they are highly dedicated to developing and improving the performance of vector search, adopting Milvus to Shotit would be beneficial in the long term. After the integration, Shotit does boost its search performance significantly at about 100x speed compared with the original Apache Solr under the same twenty million scale dataset. Applying it to the Blender Open Movie dataset and a 50 million scale proprietary TV genre dataset still performs excellently, retrieving satisfactory results within seconds. More detail about the integration will be discussed in the following THEORY REASONING section.

\section{THEORY REASONING}
In this section we will elaborate on two typical approaches to image-to-video search, the CNN approach and the LireSolr approach proposed by trace.moe. Pros and cons of them are discussed and the optimized approach that Shotit takes is introduced.

\textbf{CNN approach}. The rough system framework of the CNN approach is illustrated in Figure.5, which is utilized in the two reference papers\cite{CNNI2V1024}\cite{CNNI2V4096}. Users provide an image to the system, and then the system processes the image and extracts it to a high dimensional vector using a deep learning CNN model excluding the last softmax layer in charge of classification. After that, the vector is sent to a distributed vector similarity comparison service to match it to enormous pre-processed image vectors by Euclidean distance or cosine distance, etc, in an in-memory computing manner.

Pros: abundant, fast, and good. Because of the blossom of ML/DL, the CNN approach is nearly a de facto destination to implement image-to-image search solutions. Plenty of production-grade machine learning frameworks are well-developed and popularized, e.g., Pytorch\cite{pytorch01}, Tensorflow\cite{tensorflow01}, and PaddlePaddle\cite{paddlepaddle01}. When it comes to a specific technology stack to implement, the index procedure can use Keras\cite{keras01}, Fastai\cite{fastai01}, Towhee\cite{towhee01}, etc., And the search procedure can be powered by Faiss\cite{faiss01}, Milvus\cite{milvus01}, Vsearch\cite{vsearch01}, OpenDistro\cite{opendistro01}, Vald\cite{vald01}, Pinecone\cite{pinecone01}, Qdrant\cite{qdrant01}, etc. The emerging development of vector database provides excellent search performance and the machine learning community is mature and supportive. Such a combination is attractive for commercial use.

Cons: While the search procedure is able to run only in CPU, deploy and scale in a cloud-native environment, the index procedure of the CNN approach mostly involves expensive GPU-enabled hardware. Besides, as mentioned before, the dimensionality of feature vectors to index and search is high, usually above one thousand. Although this is suitable for million-scale search tasks, when it comes to billion-scale, it would certainly cause huge memory consumption. In addition, the byte size of machine learning frameworks is relatively costly. Finally, in the scenario of image-to-video search, an extra video processing part is needed. we were unable to find a full-fledged open source implementation utilizing this approach to inspect and dig in. Just literature reference is not enough and pragmatic for our ambition to provide an efficient method and implementation.

\begin{figure}[h]
  \centering
  \includegraphics[width=\linewidth]{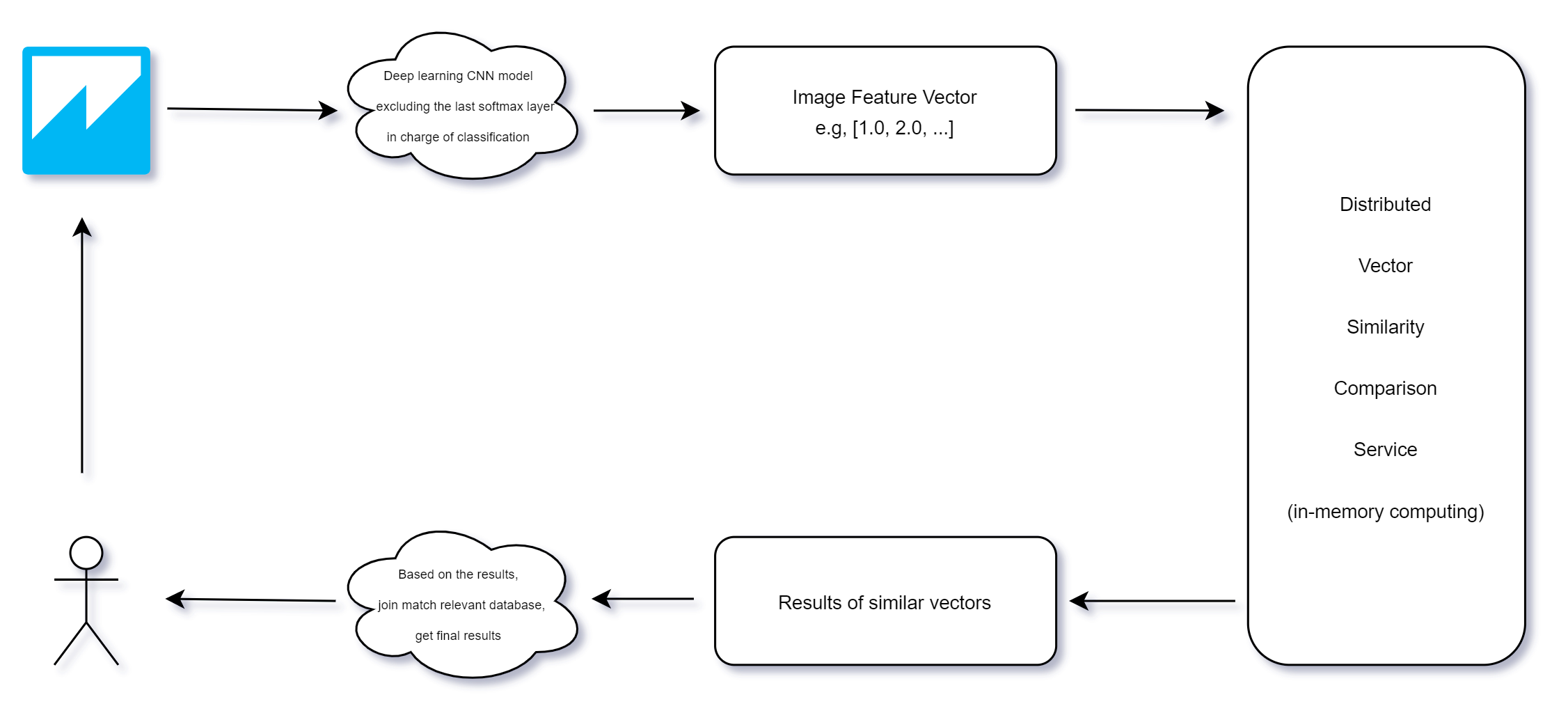}
  \caption{The CNN approach's structure}
  \Description{The CNN approach's structure}
\end{figure}

\textbf{LireSolr approach}. The LireSolr approach that trace.moe takes is in this way, as Figure.6 shows. The image that users provide is handled by the LireRequestHandler class or ParallelSolrIndexer class of LireSolr, generating an image feature hash string utilizing Color Layout\cite{colorlayout01}. Then the image feature hash string is conveyed to Apache Solr, where it is compared with other hash strings stored at disk, being loaded into memory by JVM parallelly and continuously to compare and match with cosine distance under the hood. Since Apache Solr is an inverted index keyword search engine relying on JVM, the author of trace.moe splits up the hash data into 32 solr cores and performs search concurrently with one high-end multi-core machine to make it search-performant, as mentioned before.

\begin{figure}[h]
  \centering
  \includegraphics[width=\linewidth]{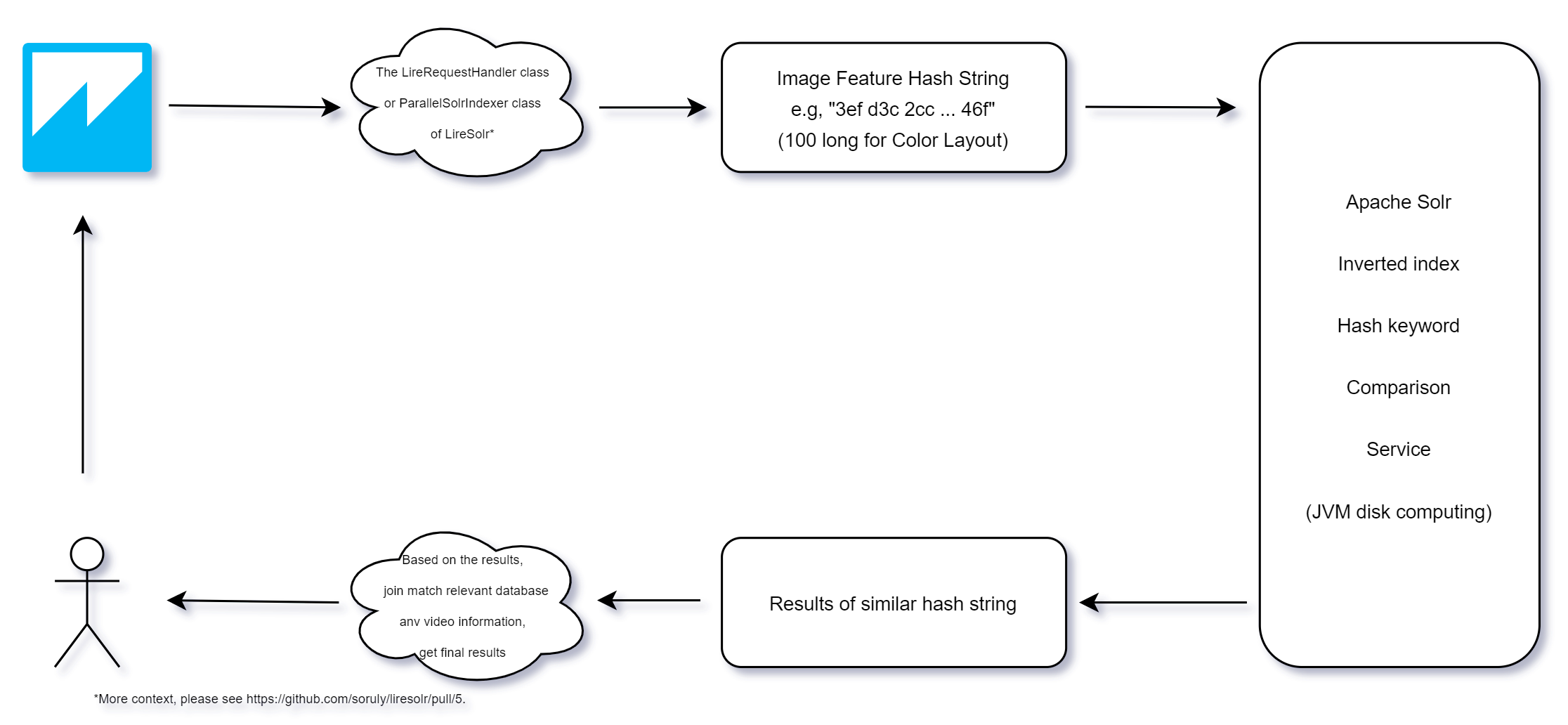}
  \caption{The LireSolr approach's structure}
  \Description{The LireSolr approach's structure}
\end{figure}

\begin{figure}[h]
  \centering
  \includegraphics[width=\linewidth]{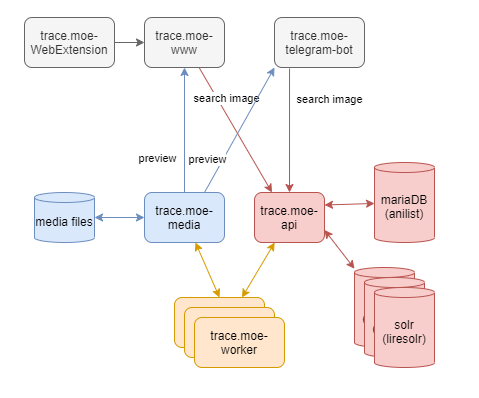}
  \caption{The system design of trace.moe}
  \Description{The system design of trace.moe}
\end{figure}

\begin{figure}[h]
  \centering
  \includegraphics[width=\linewidth]{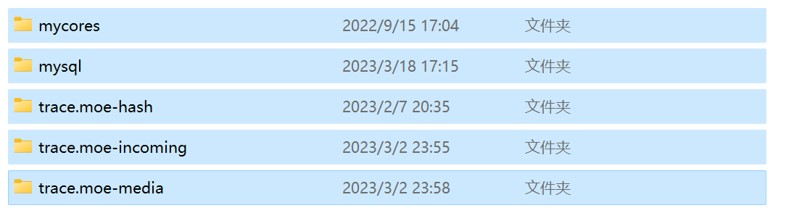}
  \caption{The five folders that trace.moe operates}
  \Description{The five folders that trace.moe operates}
\end{figure}

Pros: abundant, fairly fast, and good. The simplicity and effectiveness of the Color Layout descriptor to resolve image-to-video search impress us a lot. Due to the high-end multi-core machine, its search speed is decent. Apache Solr and LireSolr themselves are smaller runtimes, 10x smaller in byte size than machine learning frameworks. The latest version of trace.moe is written in JavaScript, containerized, and open-sourced, enabling us who are skilled in JavaScript to dig in conveniently. And considering its full-fledged functionality of image-to-video search in its particular genre - anime, with full video processing support and years-long search optimization, we are greatly inspired to adopt trace.moe to make it genre-neural and further improved. 

Cons: The downside of the LireSolr approach that trace.moe adopts locates in the search procedure. The disk-first mechanism of JVM that Apache Solr provides is not capable of tackling large-scale dataset search under low-compute limitations, as discussed before in INTRODUCTION. Besides, after completely understanding the system design of trace.moe, illustrated in Figure.7 sourced from \cite{tracemoe02}, we discovered that trace.moe is purely designed to operate upon a local file system, namely with compute unit and storage unit bundled together.

The full steps of trace.moe's index procedure are in this way. Five folders need to be created first, mycores, mysql, trace.moe-hash, trace.moe-incoming, and trace.moe-media. The administrator of trace.moe or auto-crawler places video files under the trace.moe-incoming folder, and then trace.moe-worker-watcher notices the change of the trace.moe-incoming folder. Trace.moe-worker-watcher will upload those video files to the trace.moe-media folder that the trace.moe media server is in charge of. After the upload, the video files in the trace.moe-incoming folder will be deleted, and the trace.moe-api server will maintain a SQL table to insert new records about the video files and keep maintaining for state management. Trace.moe-worker-hasher then will be notified to use LireSolr to generate hash data to the trace.moe-hash folder, as compressed XML files. Finishing hashing, trace.moe-worker-hasher will send its feedback to the trace.moe-api server, and next trace.moe-api will update the state and notify trace.moe-worker-loader, guiding it to load the hash data to the solr cores inside the mycores folder, according to Least Populated Core or Rounding-Robin. When these steps are done, the trace.moe-api server is ready to handle search requests. Receiving search request, trace.moe-api server will send the query image to LireSolr for hash generation, then get results from Apache Solr and assemble them to send the final results as a responce. The whole steps are centered in the disk-based file system, literally the five folders. 

Although these disk-centered operations are proper for the high-end multi-core machine of trace.moe, it is not elastic to deploy in a cloud-native manner. The exponentially increased video frame dataset we experimented with disclosed the unavailability of trace.moe in our situation. The video files being accumulated massively, storing them at disk is unproper because cloud vendors only provide limited-volume disk for a cloud server, and purchasing extra disk is expensive. On the other hand, object storage is much more affordable. Cloud vendors like AWS permit users to access object storage S3 from EC2 freely and swiftly\cite{aws01}. Refactoring trace.moe-media to be object storage adaptive is a promising direction.

\textbf{How Shotit adopts}. Turning back to Shotit, considering the high-end multi-core machine is hard for us to satisfy, we inspected the source code of trace.moe and realized the step that LireSolr takes to convert images to hash strings is vector-aware\cite{liresolr02}, cosine distance under the hood. On account of the unavailability of searching in Apache Solr in our situation, the promising vector database Milvus is an alternative to experiment to boost search performance. In the end, the approach Shotit utilizes to resolve image-to-video search is illustrated in Figure.9. In the beginning, the same as the LireSolr approach, users pass a target image to Shotit to be handled by the LireRequestHandler class or ParallelSolrIndexer class of LireSolr, generating an image feature hash string utilizing Color Layout. With a customized utility written in JavaScript, the image feature hash string is converted into a vector, from 100 keywords to a 100-dimensional vector. After that the vector is sent to the vector database Milvus, comparing it with other vectors preloaded in memory. Soon the results of similar vectors are returned quickly, being assembled with other related video information as final results to the users.

\begin{figure}[h]
  \centering
  \includegraphics[width=\linewidth]{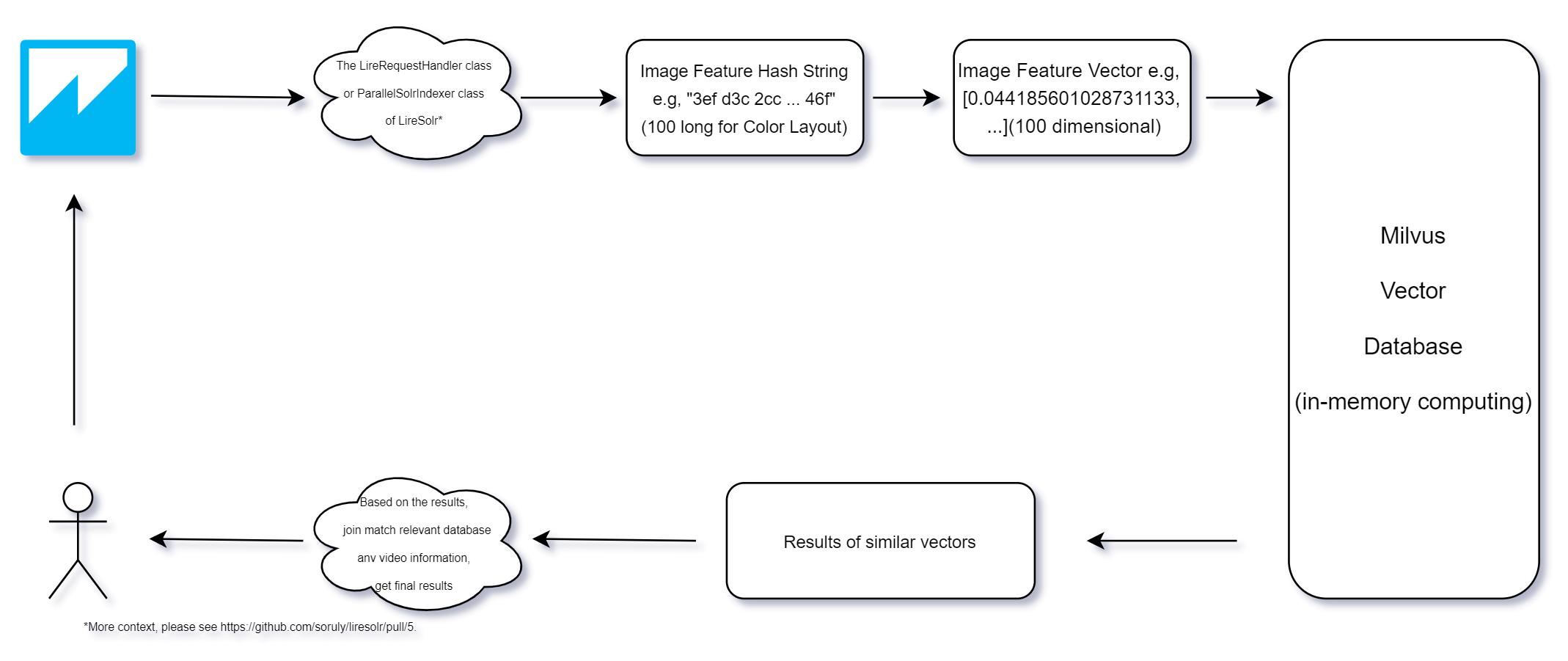}
  \caption{The Shotit approach's structure}
  \Description{The Shotit approach's structure}
\end{figure}

From the experiments we performed at a 4 Core 32GB RAM Intel Xeon Gold 6271C cloud machine with two datasets (see Table.2), a 50 thousand-scale Blender Open Movie dataset, and a 50 million-scale proprietary TV genre dataset, such an approach receives a significant boost in search performance. For readers who are interested in having an intuitive experience of our experiments, you may refer to the Shotit Demo site regarding the Blender Open Movie dataset, \url{https://shotit.github.io}.

\begin{table}
  \caption{Shotit search performance}
  % \label{tab:freq}
  \begin{tabular}{ccl}
    \toprule
    Dataset & Vector volume & Search time\\
    \midrule
    Blender Open Movie & 55,677 & within 5s\\
    Proprietary TV genre dataset & 53,339,309 & within 5s\\
  \bottomrule
\end{tabular}
\end{table}

\section{ARCHITECTURE DESIGN}
To better illustrate the architecture of Shotit, the following Figure.10 provides a detailed big picture about index and search.
\begin{figure}[h]
  \centering
  \includegraphics[width=\linewidth]{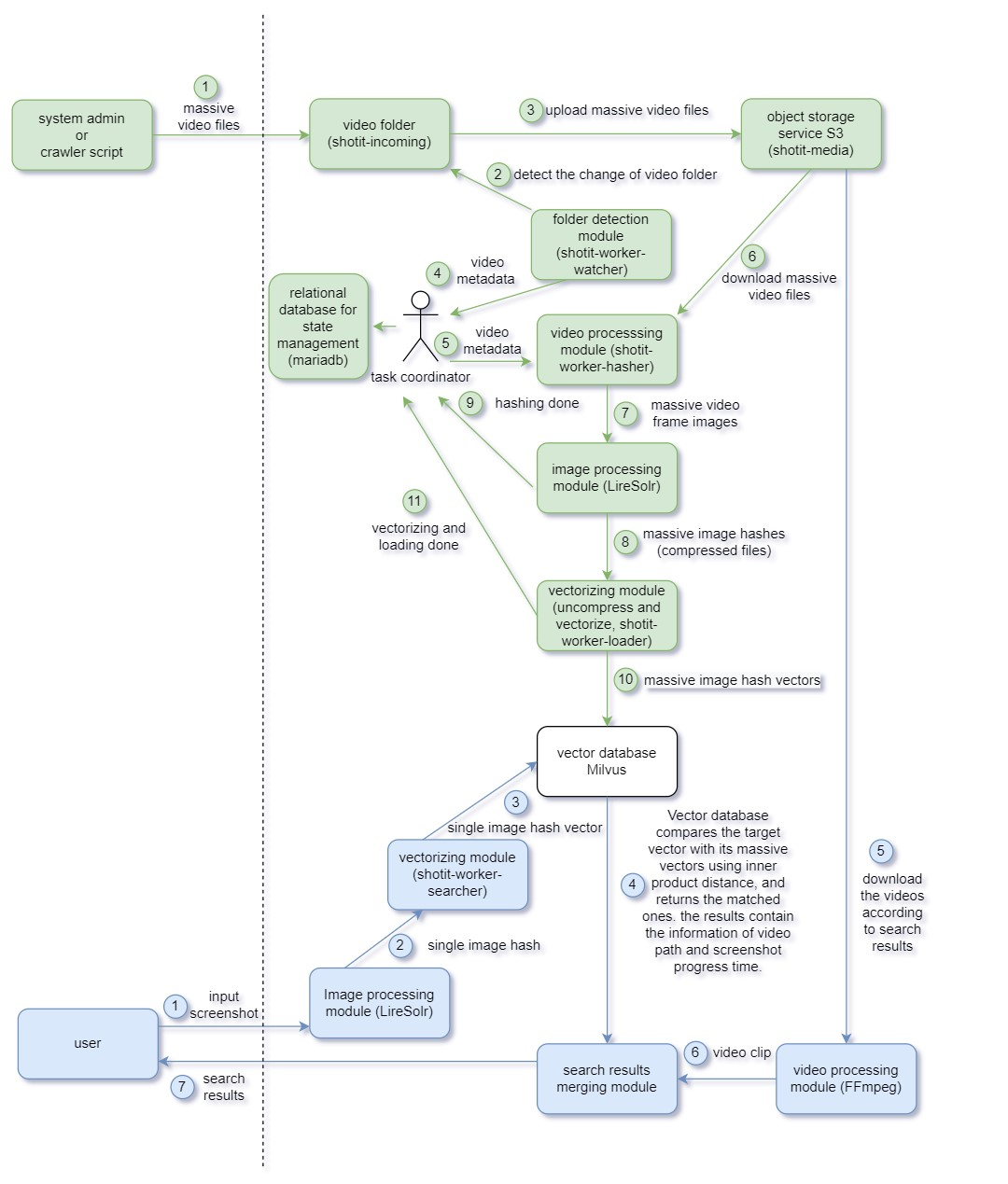}
  \caption{The big picture of Shotit}
  \Description{The big picture of Shotit}
\end{figure}

\textbf{Index:}
\begin{itemize}
\item {\texttt{Step 1}}: A system administrator or crawler script uploads massive video files to the video folder named shotit-incoming that shotit-worker-watcher keeps detecting.
\item {\texttt{Step 2}}: The folder detection module shotit-worker-watcher detects the change of video folder.
\item {\texttt{Step 3}}: Shotit-worker-watcher uploads the massive video files to an object storage service delegated by shotit-media. As for the object storage service, it is powered by the open-sourced object storage implementation MinIO\cite{minio01} and can be configured locally at disk or remotely provided by other cloud object storage services.
\item {\texttt{Step 4}}: Shotit-worker-watcher sends the signal of video metadata to the task coordinator shotit-api which is in charge of maintaining a relational database for state management and tracking. Shotit-api will create a new record at the relational database MariaDB\cite{mariadb01} and assign its state as UPLOADED.
\item {\texttt{Step 5}}: Shotit-api sends the signal of video metadata to the video processing module shotit-worker-hasher.
\item {\texttt{Step 6}}: Shotit-worker-hasher downloads massive video files in whole according to the signal of video metadata and shotit-api assigns the respective record's state as HASHING.
\item {\texttt{Step 7}}: Shotit-worker-hasher utilizes ffmpeg\cite{ffmpeg01} to process the massive video files to generate massive video frame images.
\item {\texttt{Step 8}}: Inside shotit-worker-hasher, the image processing module powered by LireSolr harvests massive image hashes as compressed XML files by utilizing the Color Layout descriptor.
\item {\texttt{Step 9}}: Shotit-worker-hasher reports to shotit-api that the hashing process is done and shotit-api assigns the respective record's state as HASHED.
\item {\texttt{Step 10}}: The vectorizing module shotit-worker-loader un-compresses the XML files, reads them as a hash list, de-duplicates the same hashes to only one within every 2 seconds, vectorizes them from hash strings to hash vectors, and then loads them into the vector database Milvus. Meanwhile, shotit-api assigns the respective record's state as LOADING.
\item {\texttt{Step 11}}: Shotit-worker-loader notices shotit-api that the vectorizing and loading operations are done. Shotit-api then assigns the respective record's state as LOADED. The index procedure completes.
\end{itemize}

\textbf{Search:}
\begin{itemize}
\item {\texttt{Step 1}}: The user passes an input image to the image processing module driven by LireSolr.
\item {\texttt{Step 2}}: Within the image processing module, LireSolr transforms the image into a single image hash string.
\item {\texttt{Step 3}}: Shotit-worker-searcher, the vectorizing module, converts the single image hash string to a single image hash vector.
\item {\texttt{Step 4}}: The vector database Milvus compares the target vector with its massive vectors using inner product distance (cosine distance), and returns the matched ones. The results contain information on the video path and the progress time of the exact frame.
\item {\texttt{Step 5}}: The video processing module downloads videos according to the search results when the user visits the multimedia links in the search results response. When the object storage service is remote, the videos are only composed of necessary HLS files, otherwise they are the whole mp4 files.
\item {\texttt{Step 6}}: The video processing module utilizes ffmpeg to generate video clips for the user.
\item {\texttt{Step 7}}: The search results merging module places the results from the vector database and the respective image link and video clip link together, sending the final results as a JSON response to the user.
\end{itemize}

\section{NOTABLE OPTIMIZATION POINTS}
\textbf{Vectorizing accuracy}. At the stage of converting image hash string to image hash vector, since we are using JavaScript, one choice is to use the number primitive of JavaScript which follows the IEEE Standard 754\cite{1985--ieee754} that represents value in a double-precision 64-bit binary format. While the number primitive fits most use cases in JavaScript, in our image-to-video scenario the precision is not enough, and higher-rational precision is needed. The jsbi-calculator
code snippet below at APPENDIX is our core JavaScript code to
perform this task.

The idea is this way. First the function receives an image hash string generated by LireSolr, e.g., "3ef d3c 2cc 7b6 9dd 2b6 549 852 582 dfd c5e c01 6af ccf 46f 1a5 5b 4a6 f8b 6d2 6a9 48d 2a1 59d ed5 b78 ac3 75 44d c15 cb3 954 1d9 44f 3a3 15b 44d 331 603 43d fb ef1 4e7 46 e92 ec6 848 c7c 8e8 8df 441 39a aa 6d6 911 9f9 d6f c2c 942 3b3 5b2 94c 521 a4c 6ac b38 7a9 584 d2a 5e3 c30 da1 733 12c fc3 dbd 152 3fa 15a b81 c24 cb beb e21 357 a0e 48e 300 19 827 2c6 b67 651 dba 9a4 b4b 85 d75 f78 c30". These words are expressed in base 16 format, so we convert them back to base 10, getting "[1007, 3388, 716, 1974, 2525, 694, 1353, 2130, 1410, 3581, 3166, 3073, 1711, 3279, 1135, 421, 91, 1190, 3979, 1746, 1705, 1165, 673, 1437, 3797, 2936, 2755, 117, 1101, 3093, 3251, 2388, 473, 1103, 931, 347, 1101, 817, 1539, 1085, 251, 3825, 1255, 70, 3730, 3782, 2120, 3196, 2280, 2271, 1089, 922, 170, 1750, 2321, 2553, 3439, 3116, 2370, 947, 1458, 2380, 1313, 2636, 1708, 2872, 1961, 1412, 3370, 1507, 3120, 3489, 1843, 300, 4035, 3517, 338, 1018, 346, 2945, 3108, 203, 3051, 3617, 855, 2574, 1166, 768, 25, 2087, 710, 2919, 1617, 3514, 2468, 2891, 133, 3445, 3960, 3120]". Because the vector database Milvus requires vector normalization before insertion, the square root operation is performed next and then each element inside the vector gets normalized.

As discussed before, the number primitive-based Math.sqrt operation of JavaScript is not precise enough, here we adopt an NPM library jsbi-calculator\cite{jbc01} developed by us to tackle this problem. Able to perform arbitrary (up to 18 decimals) arithmetic computation as well as square root operation, jsbi-calculator implements BigDecimal-based arithmetic operations in JavaScript and uses reverse polish notation\cite{rpn01} to wrap up to provide an easy interface to use. Noticeably, the incubation of this NPM library is inspired by GoogleChromeLabs' jsbi project, and it has made a significant contribution to jsbi by firing an issue and PR\cite{jsbi02} to it. 

After all these steps, the generated image hash string is this way, "[0.044185601028731133, 
0.1486601949208948, 0.031416971535820744, 0.0866160639828354, 0.11079309096082036, ... , 0.10829201920447709, 0.1268526043436561, 
0.0058358340981343, 0.1511612666772381, 0.17375866938805887, 0.13690076982089486]", with 18 decimals-long high 
precision.

\textbf{Border cut}. To increase the search correctness, the target image used to search can be optimized. Mostly the target image shot from a video contains black borders. OpenCV implements a function findContours\cite{opencv02} which can be used to facilitate cutting black borders. The author of trace.moe takes advantage of this function and provides an optional parameter to cut the black borders of the target image before vector comparison. From the report\cite{tracemoe2018} provided by the author in 2019, the sample image could achieve 96.3\% similarity, compared to 89.4\% before without border cut.

\textbf{Video clip scene detection}. When returning the video preview, the video preview should be retained in the same scene. If not doing so but using a fixed time offset instead, some unfriendly users might make use of this to search one more time to fetch the previous or next scene of the original video. Once this operation gets repeated, the users might be able to read the whole video, which might cause copyright issues. With this in mind, the author of trace.moe utilizes a special technique\cite{tracemoe2018} to ensure the video clip is retained in the same scene. First given a relatively long-time range, use ffmpeg to split and generate all the frame images, then for each frame image, add up all the pixel numbers to get a sum. Next, find from the middle, and move forward or backward until a rapid change of the sum happens. Mark that time offset as the final result to get a video clip. According to the author's words, its accuracy could reach 87\% but the methodology is agnostic to us.

\section{PERFORMANCE BENCHMARKS}
Previously table.2 shows our achievement after refactoring to utilize the vector database Milvus to speed up search. Referring to the prestigious CS textbook, {\itshape Computer Systems: A Programmer's Perspective}\cite{CS01}, we found Amdahl's Law explained there in Chapter 1 provides us a clear insight into how to further improve the performance of Shotit. The idea can be conceived as that one part of a system is sped up, the impact on the thorough system performance relies on both how significant this part was and how much it sped up. Suppose it costs $T_{old}$ for a system to execute a certain program. And one part of the system takes a fraction $\alpha$ of this time. If we speed up the performance to k times smaller, namely $\alpha T_{old}/k$, then the execution time can be calculated as 
$$
T_{new} = (1 - \alpha)T_{old} + (\alpha T_{old})/k \\ 
= T_{old}[(1 - \alpha) + \alpha/k]
$$

From this formula, we can get the speedup S = $T_{old}/T_{new}$ as
$$
S = \frac{1}{(1 - \alpha) + \alpha/k}
$$

Consider an example where $\alpha$ is 0.6 and k is 3. Then the speedup is 1/(0.4 + 0.6 / 3) = 1.67x. As you can observe, a 3-times smaller improvement to one part can only lead to a significant 1.67x improvement to the whole system. From our experiments, Shotit achieves a 100x speedup compared with the original Apache Solr under the same twenty million scale dataset after adopting the vector database Milvus, from around 100s to only about 1s. Hence we can infer that the k here is really substantial. Shotit is composed of two parts, the index part, and the search part. The critical search performance gets improved significantly now. The index building performance ought to be improved too. Although only a limited number of people are involved in the index part when running Shotit in production, improving the index building time will significantly make Shotit much more ease of use for potential developers considering that Shotit is publicly available in the open source space. The index building performance is a direction we aim at.

\section{CONCLUSION \& FUTURE WORKS}
In this paper we present our image-to-video search engine Shotit, review some research backgrounds about image-to-video search, cloud-native technologies, music retrieval, peer-to-peer file sharing, and approximate nearest neighbor search, reason why Shotit is compute-efficient by analyzing two typical approaches to image-to-video search, the CNN approach and the LireSolr approach. What followed is the detailed big picture of Shotit's index and search. Finally, we reveal some notable optimization points and demonstrate Shotit's performance progress from a numerical point of view.

In the future, many works are remaining to be done. First, because Shotit is open-sourced with the Apache II license at GitHub, documentation and translation are needed to drive its adoption. Second, according to our experiments with the two datasets, the index building experience of Shotit is a bit tedious and time-consuming, in contrast with the swift search experience. Significant engineering effort is needed to investigate this problem. Third, periodic dependencies updates and tag releases are necessary to prevent security threats. Fourth, considering LireSolr has twelve different image descriptors\cite{liresolr01} and only Color Layout is leveraged, other image descriptors may perform better under some aspects since obviously Color Layout is not that capable to cope with dark images. A comprehensive numerical analysis of how these image descriptors perform under the scenario of Shotit is worth investigating. Last but not least, as the backbone Milvus is developing rapidly and keep bolstering its performance, Shotit needs to keep up with the upstream update to benefit from open source. 

Honestly, this paper apparently has some drawbacks. We merely present a method and implementation to resolve the image-to-video search problem. Yet the effectiveness of its compute-efficient property is less proven numerically, only with empirical performance in about two datasets. For those readers who are skeptical about numerical deduction, please recognize the effort and limitations of this paper. 

\section{ACKNOWLEDGEMENTS}
Great credits to the author of trace.moe soruly, whose perseverance towards the development of trace.moe since 2015 and dedicated action to migrate the codebase from PHP to JavaScript provides us a significant opportunity to incubate a genre-neutral image-to-video search engine.

To Dr. Mathias Lux, the author of LireSolr.

To Mr. Basker George from Shenzhen University, whose encouragement and paper polish work help a lot.

To Prof. Peng Xiaogang from Shenzhen University, whose recognition of contributing Shotit to the multimedia research field initiated the birth of this paper.

To Chen Jiaxian, a Ph.D. student at Shenzhen University, whose literature review expertise guided us to a clear position on which research domain Shotit belongs to.

To He Wangqian, a Ph.D. student at Shenzhen University, whose first impression about the paper draft realized us to add more intuitive diagrams.

\bibliographystyle{ACM-Reference-Format}
\bibliography{shotit-base}

\appendix
\section{jsbi-calculator code snippet}
\begin{lstlisting}
import JBC from 'jsbi-calculator';
const { calculator, BigDecimal } = JBC;

/**
 * getNormalizedCharCodesVector
 * @param {String} str
 * eg. '3ef d3c 2cc ... d75 f78 c30'
 * @param {Number} length
 * @param {Number} base
 * @returns []Number
 */
const getNormalizedCharCodesVector = (str, length = 100, base = 1) => {
  const arr = str.split(" ").map((el) => parseInt(el, 16));
  let charCodeArr = Array(length).fill(0);

  // arr.length should be less than parameter length
  for (let i = 0; i < arr.length; i++) {
    let code = arr[i];
    charCodeArr[i] = parseFloat(code / base);
  }

  const norm = BigDecimal.sqrt(
    String(
      charCodeArr.reduce((acc, cur) => {
        return acc + cur * cur;
      }, 0)
    )
  ).toString();

  return charCodeArr.map((el) => parseFloat(calculator(`${el} / ${norm}`)));
};
\end{lstlisting}
\end{sloppypar}
\end{document}